\documentstyle[onecolumn]{mn}

\title{Quantifying the topology of
large--scale structure}
\author[Peter Coles, Andrew Davies \& Russell Pearson]{Peter 
Coles, Andrew Davies \& Russell C. Pearson,\\
Astronomy Unit, School of Mathematical Sciences,
Queen Mary and Westfield College, Mile End Road, London E1 4NS}
\begin{document}

\maketitle

\begin{abstract}
We propose and investigate a new algorithm for quantifying the
topological properties of cosmological density fluctuations.
We first motivate this algorithm by drawing a formal distinction
between two definitions of relevant
topological characteristics, based
on concepts, on the one hand, from differential topology and,
on the other, from integral geometry. The former approach
leads one to concentrate on properties of the contour surfaces
which, in turn, leads to the algorithms CONTOUR2D and CONTOUR3D
familiar to cosmologists. The other approach, which we adopt here,
actually leads to  much simpler algorithms in both two and
three dimensions. (The 2D algorithm has already been introduced to
 the astronomical literature.)
We discuss the 3D case in some detail and compare results
obtained with it to analogous results using the CONTOUR3D
algorithm.
\end{abstract}


\section{Introduction}

The recent dramatic increases in the quality and quantity of
galaxy redshift data available to cosmologists, mapping larger and larger
regions of our local Universe with increasing precision, 
has stimulated considerable interest in the development of sophisticated
analysis techniques capable of extracting the essential characteristics
of the spatial pattern displayed by such data. To get the most out of
the current and next generation of redshift surveys requires techniques
that can go beyond the traditional approach based on correlation functions
(Peebles 1980). Only with appropriately specialised tools can such
redshift data be used most effectively 
to constrain theoretical models of structure formation.

One approach towards characterising the properties of the large-scale
distribution of matter in the Universe has been to measure
objectively certain {\em topological} characteristics of the large-scale density
field. This approach has generally involved the construction of contour
surfaces of equal matter density. Then, with an appropriate algorithm,
topological properties such as the {\em genus} (which quantifies the
connectivity) of these
surfaces can be calculated and compared with theoretical calculations
or N--body simulations. This idea has led to a number of detailed
analyses of both three--dimensional data (Gott, Melott \& Dickinson 1986;
Hamilton, Gott \& Weinberg 1986; Weinberg, Gott \& Melott 1987;
Melott, Weinberg \& Gott 1988; Gott et. al. 1989;
Melott 1990) and angular (projected)
distributions (Melott et al. 1989; Gott et al. 1992). A good review
of this kind of analysis can be found in Melott (1990), where FORTRAN
implementations of the algorithms normally used (i.e. 
CONTOUR2D \& CONTOUR3D) are also given.

A particular advantage of the ``genus'' statistic, as it has come to be
known, is that the mean value of this quantity can be calculated exactly
for a Gaussian random field (Doroshkevich 1970;
Adler 1981; Bardeen et al. 1986; Tomita 1986), 
the generic assumption for the initial
density field in gravitational instability models. The behaviour of the
genus can therefore be used to 
constrain the possibilities for non--Gaussian
initial fluctuations using observations of galaxy
clustering. The availability of the mathematical machinery
to derive topological characteristics exactly for random density fields is
another one of the motivations for using these characteristics as
statistical descriptors.

In this paper we shall examine the  ``topological'' approach 
from a mathematical point of view. We shall show that there are
two distinct routes leading to technically different definitions
of the relevant topological characteristics but that, in situations
of relevance to cosmology, both the characteristics one derives are
numerically equivalent or, at least, nearly so. 
The first route, based on a branch of
mathematics known as differential topology, provides an
elegant route to the derivation of the analytical results mentioned
in the previous paragraph, through properties of the spatial
derivatives of the contour surfaces. The algorithms CONTOUR2D
and CONTOUR3D, which are fairly complex, are also based on this
line of thought. The second approach is based on a different
mathematical foundation, provided by integral geometry (IG) rather than
differential topology (DT). Unlike the first, it does not facilitate
the elegant derivation of analytic results for Gaussian random
fields. On the other hand, it does lead to algorithms for extracting
topological descriptors from a given data set which are far simpler
than those emerging from the former route.

Our approach is first to outline (briefly) these two mathematical
approaches and explain how they lead to computations of essentially
the same characteristics, but by very different routes. We then
explain an algorithm based on the IG approach and show that it gives
results which are indistinguishable from those obtained using
CONTOUR3D, but which are obtained at a fraction of the computational cost.
Those readers not interested in the mathematical niceties can skip
directly to Section 3, where we describe our proposed algorithm and
show the results of a test. 

\section{Technicalities}
We shall be discussing the properties of {\em excursion sets} of
a random field, $X({\bf r})$, defined on an $n$--dimensional space, 
which we assume to be equipped with a Cartesian coordinate system
made up of the unit vectors ${\bf e}_1,\ldots, {\bf e}_n$,
so that ${\bf r}\in \Re^n$. The excursion set of $X$, called ${\cal A}_u$
is that subset of the space where $X$ exceeds some specific level, say $u$. 
In general, the excursion sets of a random field may consist of
many disjoint pieces, each of which may be simply or multiply connected.
In cosmological applications, $X$ is usually the density contrast
$X=\delta\equiv(\rho-\rho_0)/\rho_0$, where $\rho_0$ is the mean matter density;
$\delta$ has zero mean and, we assume, variance $\sigma^{2}$.

In practical situations
 the cosmological mass distribution is sampled, discretely, by
some form of cosmic object (optical galaxy, infra-red
galaxy, cluster of galaxies, etc.). In an N--body experiment,
one is similarly hampered by discreteness effects.
In order to construct the random field $X({\bf r})$ one therefore needs
to smooth
the discrete distribution by convolving with an appropriate
filter (Melott \& Dominik 1993). Normally, this is done so as
the produce a field defined on a Cartesian lattice, the most convenient
representation for the forthcoming analysis.

We also have to consider the possibility that the excursion set may
have an external boundary imposed upon it. For any particular sample,
this will be determined by the edge of the survey region. In a simulation,
the boundary may be the edges of the (usually cubic) region that is being
simulated. Alternatively, in the second case, one can adopt periodic
boundary conditions which dispense with the need for such considerations:
here the space will be finite but will have no boundary. We shall touch
on these considerations later.

The mathematical literature describes two distinct approaches towards
characterising the topological properties of such sets. For elementary
background material on related subjects, see Nash \& Sen (1983). Much
of the material presented here can be found in more rigorous fashion
in Adler (1981).

\subsection{Differential Topology}
The first approach we shall discuss here employs concepts from the
field of differential topology. We describe this approach first, as
it is the more familiar in the astronomical literature.
Essentially, one concentrates
on properties of the bounding
surfaces of the excursion set $\partial{\cal A}_u$
rather than on the set itself. Specifically, one relates topological
properties of the excursion set to properties connected with the
curvature of the surfaces bounding this set. This requires at the
outset a number of non--trivial technical conditions to hold, relating to 
differentiability of these surfaces, which we shall not discuss here;
see Adler (1981) for a technical discussion.

We start in a general way by considering a generic
two--dimensional manifold ${\cal M}$, with a 
(one--dimensional) boundary
$\partial {\cal M}$, which is piecewise smooth (i.e. it may have vertices
where it is not differentiable). The general form of the Gauss--Bonnet
theorem states that
\begin{equation}
\sum_{i=1}^{n} (\pi-\alpha_i) + \int_{\partial {\cal M}} k_g ds
+\int_{\cal M} k dA =  2\pi \chi_{_E}({\cal M}),\label{eq:ep1}
\end{equation}
where  $\alpha_i,\ldots \alpha_n$ are the $n$ interior angles of the
vertices of the boundary $\partial {\cal M}$, $k_g$ is the  geodesic
curvature of the curve $\partial {\cal M}$, and $k$ is the ordinary
(Gaussian) curvature of the manifold ${\cal M}$; $ds$ and $dA$ are
elements of length and area respectively. The quantity $\chi_{_E}$
is called the Euler, or Euler--Poincar\'{e}, characteristic of the
manifold.

The case where there are vertices in the boundary is not particularly
relevant in cosmology, so we shall drop this possibility from now
on. The theorem (\ref{eq:ep1}) is relevant to the topology of
both two-- and three--dimensional excursion sets. First consider
a two dimensional excursion set defined on a flat plane.
Here the Gaussian curvature is everywhere zero and the Euler
characteristic is simply given by an integral of the
line curvature around the boundaries of the excursion set:
\begin{equation}
2\pi \chi_{_E} = \int k_g ds. \label{eq:ep2}
\end{equation}
In two dimensions, $\chi_{_E}$ is simply the number of isolated
regions minus the number of holes in such regions.

On the other hand, if we have a three--dimensional excursion set
bounded by a two--dimensional surface which itself has no boundary:
(e.g. a sphere, torus, etc) then the Euler characteristic is
simply the integral of the Gaussian curvature of the
surface over all compact pieces of the excursion set:
\begin{equation}
2\pi \chi_{_E} = \int k dA.
\label{eq:ep3}
\end{equation}
In the three--dimensional case it has become fashionable to
refer instead to the {\em genus}, $g$, an alternative topological
quantity which is, roughly speaking, defined as the number
of ``handles'' a compact two--dimensional surface possesses. A sphere
has zero genus, a torus has unit genus, and so on. It is straightforward
to show (Nash \& Sen 1983) that 
\begin{equation}
\chi_{_E}=2(1-g).
\label{eq:ep4}
\end{equation}
The genus applies to three--dimensional sets (with two--dimensional
surfaces) and has no direct equivalent in the case of
two--dimensional sets (with one--dimensional ``surfaces'');
the Euler characteristic is, however, equally well--defined in
both cases. 

Notice that the genus (defined in this way) is not an additive
characteristic. One can see this by taking the case of a
set made up of two disjoint
spheres. Each of these has genus zero so one would imagine the
total genus of the set to be zero. On the other hand, the
integrated curvature around each sphere must be $4\pi$ so the
total for the set is $8\pi$. 
The Gauss--Bonnet theorem (\ref{eq:ep3}) then
tells us that $\chi_{_E}=2$ for each sphere and therefore
the total $\chi_{_E}=4$. Straightforwardly applying
eq. (\ref{eq:ep4}) then suggests $g=-1$ for this set.
For the reason, and for technical reasons connected with
sets which intersect boundaries (see Sec. 3 below),
the practical implementation of the
``genus statistic'' in CONTOUR3D in fact does not employ
the genus itself, but defines a quantity
\begin{equation}
g_S=-\chi_{_E}/2,
\end{equation}
which is additive. It is $g_S$ rather than $g$ as defined by equation
(\ref{eq:ep4}) which is termed the ``genus'' in most of the cosmological
literature.

One of the advantages of this approach is that one can relate the
integrated curvature over a surface to the properties of the
(discrete) set of critical points of that surface. The relevant
branch of mathematics is known as Morse theory (Morse \& Cairns 1969)
and, using it, one can show, for example, that
\begin{equation}
 \chi_{_E}= \mbox{number of maxima} + \mbox{number of minima}
-\mbox{number of saddle points},\label{eq:morse}
\end{equation}
which is actually used by Bardeen et al. (1986) as the definition
of the Euler characteristic. This equation, and/or eq. (\ref{eq:ep3}),
allows one to relate global characteristics of each piece of the
excursion set to local properties of the surface bounding it.

It is particularly relevant to note that one can exploit the properties
of surface curvature to obtain the mean value of the Euler characteristic
per unit volume (or area in 2D) for any random field $X$ for which one has
has the joint probability distribution of $X$ and its first two spatial
derivatives. In particular this can be accomplished for a Gaussian random
field (Doroshkevich 1970; Bardeen et al. 1986; Hamilton et al.
1986; Tomita 1986). In fact, the
Doroshkevich (1970) result predates the first derivations of this result
in the mathematical literature (Adler 1976; Adler \& Hasofer 1976; see
also Adler 1981). Some special cases of non--Gaussian fluctuation fields
are discussed in Coles (1988). For the particular case of relevance here, 
that of a Gaussian random field in three dimensions, it can be
shown that
\begin{equation}
\chi_{_E}(\nu)  = A \left(1-\nu^{2}\right) \exp \left(-\nu^{2}/2\right),
\label{eq:epgauss}
\end{equation}
where the argument $\nu$ in equation (\ref{eq:epgauss})
is the threshold density
contrast, $\delta_{\rm t}$, defining the excursion set, 
expressed in dimensionless form: $\delta_{\rm t} = \nu\sigma$, where
$\sigma$ is the rms value of $\delta$. The quantity $A$, which
is negative, simply
depends on the {\em coherence length} of the field which, in turn,
depends only on its 
power spectrum, so that all Gaussian random fields
have a graph of $\chi_{_E}(\nu)$ which has the same shape but an amplitude
which depends on the initial power spectrum. (Recall that the power
spectrum or, equivalently, the two--point correlation function, furnishes
a complete description of the statistical properties of a Gaussian random
field). Notice that $\chi_{_E}$ is negative for $\nu=0$, indicating
multiple connectivity ($g>1$): this is the so--called ``sponge'' topology
(Gott, Melott \& Dickinson 1986). At larger values of $|\nu|$, the
topology approaches that of simply connected isolated regions (or holes)
which become rarer and rarer as $|\nu|$ increases. As first pointed
out by Doroshkevich (1970), the value of $\chi_{_E}$ per unit
volume for large $\nu$ leads to
a useful approximation to the number density of local maxima, since
each piece of the excursion set at high thresholds tends to consist
of one isolated simply connected region containing one local maximum
of $\delta$. This situation, of isolated high-density islands surrounded
by a low-density `sea' is usually called a ``meatball'' topology while
the opposite, low density regions embedded in a high-density background
is, topologically speaking, equivalent to ``Swiss--cheese''.

The equation (\ref{eq:ep3}) can also used as a basis for extracting
a measurement of $\chi_{_E}$ from a given realisation of $\delta$,
estimated  by smoothing galaxy data or particles in an N--body experiment.
Here one must find some way of estimating the derivatives of the contour
surface, for example by fitting some form of polyhedral surface to it
(Hamilton, Gott \& Weinberg 1986). The algorithms CONTOUR2D and CONTOUR3D,
written by D. Weinberg, are implementations of this idea; FORTRAN
listings of these programs are available in Melott (1990).
 
\subsection{Integral Geometry}
The other approach, which is actually the older one historically
speaking,   uses the formalism of integral geometry
(Hadwiger 1959; Adler \& Hasofer 1976; Adler 1981). 
In this approach, one considers
the excursion set to be constructed from objects known as {\em basic} sets.
Let us define a $k$--dimensional hyperplane ${\cal E}$ to be a subset
of $\Re^{n}$ such that any $(n-k)$ of the coordinates $r_i$ are fixed; it is thus
generated by $k$ of the vectors ${\bf e}_1\ldots {\bf e}_n$.
A compact set ${\cal B}\subset \Re^n$ is a basic if the intersections of 
${\cal B}$ with all possible $k$--dimensional hyperplanes 
are simply connected (including the case with $k=n$). One then
takes the excursion set ${\cal A}$
(within a finite region of space)
to be represented as the union of a finite number $m$ of basics in such a
way that the intersection of any of these basics is itself a basic;
the formal proof that this can be done exists, but
is not trivial (Adler 1981).
Given this geometrical structure, one can then construct the
Hadwiger characteristic $\phi_{_H}({\cal A})$:
\begin{equation}
\phi_{_H}({\cal A}) \equiv \sum_{(1)} \epsilon ({\cal B}_i)
-\sum_{(2)} \epsilon ({\cal B}_{i_1} \cap {\cal B}_{i_2}) + \cdots
+(-1)^{r}\sum_{(r)} \epsilon ({\cal B}_{i_1}\cap {\cal B}_{i_2} \cdots \cap
{\cal B}_{i_r}) + (-1)^{m} \epsilon({\cal B}_{i_1}\cap {\cal B}_{i_2}
\cdots \cap {\cal B}_{i_m}),\label{eq:IG1}
\end{equation}
where $\sum_{(r)}$ means a sum over all combinations of $(i_1,\ldots, i_r)$
from the elements $(1,\ldots, r)$ and $1\leq r\leq m$; $\epsilon({\cal B})$
is the indicator function which is zero if ${\cal B}$ is the null set and
equal to unity otherwise. This functional is, in fact, one of the
Minkowski functionals (Mecke, Buchert \& Wagner 1994); it possesses
properties of invariance under translations and rotations, does not
depend on how the set is partitioned into basics and also
has the important property of additivity:
\begin{equation}
\phi_{_H}({\cal A})+\phi_{_H}({\cal B})=\phi_{_H}({\cal A}\cap {\cal B})
+\phi_{_H}({\cal A}\cup {\cal B}).\label{eq:IG2}
\end{equation}
The summation (\ref{eq:IG1}) looks quite complicated, 
but can be shown to be equivalent
to the following recursive (and much simpler) definition:
\begin{equation}
\phi_{_H}({\cal A}) = \mbox{number of disjoint intervals in } {\cal A}
\,\,\,\,\,\,\,\,\,\,\,\,\,\,\,\,\,(n=1)\label{eq:IG3}
\end{equation}
and
\begin{equation}
\phi_{_H}({\cal A}) = \sum \left\{ \phi_{_H}({\cal A} \cap {\cal E}_x) - 
\phi_{_H}( {\cal A} \cap {\cal E}_{x_-} )\right\}
\,\,\,\,\,\,\,\,\,\,\,\,\,\,\,\, (n\geq 2),
\label{eq:IG4}
\end{equation}
 where
\begin{equation}
\phi_{_H}\left({\cal A}\cap {\cal E}_{x_-}\right) = \lim_{y\rightarrow 0}
\phi_{_H}\left({\cal A} \cap {\cal E}_{x-y} \right)
\label{eq:IG5}
\end{equation}
and the summation in (\ref{eq:IG4}) is taken over real $x$ where the summand is
non--zero. Note that the disjoint intervals counted in the
one--dimensional case can be degenerate (i.e. points).
This gives a very simple recursive
algorithm for measuring the
Hadwiger characteristic of the set, particularly in two dimensions
as illustrated in Figure 1.
In this case, the only allowed hyperplanes are lines. One simply
scans a line parallel to the $x$-axis upwards through increasing
values of $y$, noting the contribution to $\phi_{_H}$ every time the
number of intervals in the intersection of the line with ${\cal A}$
changes. The appropriate sum is then
\begin{equation}
\phi_{_H} = \sum_x [N(x)-N(x_-)]=\sum_j C_j,
\label{eq:IG6}
\end{equation}
where $N(x)$ is the number of disjoint closed intervals, $N(x_-)$
is the relevant limit in eq. (\ref{eq:IG5}),
and the sum is over a finite number
of contributions $C_j$.
Figure 1 shows how this works for a simply--connected set
and for one with a hole in the middle. These figures are
self--explanatory as long as one remembers to calculate the
contributions $C_j$ using the one--side limit (\ref{eq:IG5}) 
and that new intervals may be degenerate, i.e. they may
appear first as points as
one scans the line upwards. The application of this
technique in three dimensions is more difficult to visualise, but
is quite simple to program on a computer: one scans upward through the
3D excursion set, slicing it using 2D planes, and scans each plane
with 1D lines in the manner described previously.

\begin{figure}
\vspace{10cm}
\caption{The recursive calculation of the Hadwiger characteristic
in two dimensional examples. The left figure shows a simply--connected
set with $\phi_{_H}=1$, while the right figure has one region
containing one hole: $\phi_{_H}=0$. The $y$-axis is labelled
with the number of disjoint intervals and the values of
$C_j$ at the points where the value of $N(x)$ changes.}
\end{figure}

We have tacitly assumed so far that the excursion set
does not intersect any external boundary ${\cal C}$. Of course,
the question of edge effects  does not arise if one is dealing with periodic
boundaries, such as in typical $N$--body simulations. In calculating
the DG characteristics with CONTOUR3D, for example, one simply allows
the algorithm to `wrap-around' the edges of the simulation box.

Correction for boundary effects is slightly less straightforward
in the context of the IG characteristic. If we are dealing with
a simulation box, one can allow the IG algorithm to wrap-around
as in the case mentioned in the previous paragraph. This is,
in fact, equivalent to  subtracting
 off the characteristic of the set comprising
the intersection ${\cal A} \cap {\cal C}$, where ${\cal C}$ is the
coordinate boundary of the box (for a periodic simulation this is
not really a physical boundary). For example, if one has
an excursion set embedded in a cubic volume, one subtracts from the
3D characteristic the 2D characteristics generated by the intersection
of the set with the three planes with the origin at one corner:
\begin{equation}
\chi_{_H}=\phi_{_H}({\cal A}) - \phi_{_H} ({\cal A} \cap {\cal C}_0),
\label{eq:bound}
\end{equation}
where ${\cal C}_0$ indicates that part of the boundary which has the
origin at a vertex. 
This formulation of the boundary correction is a straightforward
consequence of the definition of $\phi_{_H}$ given above, in equation
(\ref{eq:IG1}). 
Notice, however, that there is a problem with this definition:
in the case of sets which intersect the boundary, but do not have
the same number of intersections with each part of the boundary, 
this procedure gives a different result depending
on which of the coordinate axes one chooses as the $x$-axis in the
above operation; see, for example, Figure (4.4.2) of 
Adler (1981). The differential geometry characteristic,
on the other hand, is invariant with respect to transformations
of the coordinate axes in this way. If we are going to use the
IG approach as an estimator of $\chi_{_E}$, therefore, we are
immediately restricted to sets which are effectively `isotropic'.
 This effectively means that we must be dealing with a large
enough sample of the random field that it contains regions
oriented in all possible ways with respect to the coordinate
axes. This is one aspect of the need to have a `fair sample'
of the Universe in order to estimate global characteristics
from local samples. Notice that this is a stronger requirement
than the requirement that the field be {\em ergodic}. The
property of ergodicity is, roughly speaking,
that spatial averages over an infinite
domain are equivalent to averages performed over the probability
distribution. For many statistical analysis tools, 
including the IG characteristic, to be useful one actually needs
the average over a finite volume to produce a result within
some acceptable range of the average of the distribution.
Whether a given technique fulfills this particular criterion
for `usefulness' is a question that has to be answered 
on an individual basis. Ergodicity would appear to be a necessary
condition for usefulness, but not in all cases a sufficient one.

Adler (1981) advocates $\chi_{_H}$ 
(he calls it $\Gamma$) as the most
appropriate definition of an IG-equivalent to the DG characteristic
$\chi_{_E}$: an alternative, not equivalent to $\chi_{_H}$, 
has been suggested by Fava \& Santalo (1979) which also
attempts to take account of the lack of coordinate invariance
of $\chi_{_H}$. Exact correction for
boundary effects can be made using the complete set of 
Minkowski functionals (Likos et al. 1995; Schmalzing, Kerscher 
\& Buchert 1996). Notice that $\chi_{_H}$ only coincides with
$\phi_{_H}$ if there are no intersections of the set with
the boundary. In addition, if the field is periodic then
these two quantities coincide again: the boundary correction
cancels as we mentioned above. If one has a cubic volume within
a non-periodic field then adopting this boundary correction
is tantamount to assuming the field is periodic outside the
volume and this will introduce some error in $\chi_{_H}$ as
an estimator of $\chi_{_E}$. In a practical situation one
would require the boundary correction to be small in this
circumstance, otherwise one cannot claim to have a `fair sample'.

Our discussion of boundary corrections has so far been restricted
to the case of cubic boundaries. In reality, boundaries of galaxy
surveys are likely to be much more complicated than this. There
are two different ways to view this problem. First, one can try to
implement a rigorously-defined boundary correction in order to 
produce an estimate from the sample which is as close as possible to
the global average defined over the probability distribution.
In this vein, Alder (1981, p. 78) gives a concise way of extending the boundary
correction in equation (\ref{eq:bound}) to more complicated
shapes. Another possibility is to attempt to eliminate the boundary
by a weighted averaging method (Coles 1988; Coles \& Plionis 1991;
Davies \& Coles 1993; Davies 1994). On the other hand, a more
pragmatic approach would suggest not worrying too much about the
accuracy of the estimate of global quantities from finite samples,
particularly if the boundary is complicated and the sample is small.
If one is really interested in testing
observations against model simulations constructed with the same
boundaries and selection effects then, as long as one does the same
correction for the data as is done for the models, 
the precision of the correction will not affect the robustness
of the test. If the correction is large compared with the
true signal, however, then such a test will have no power
as differences between simulation and data will be masked
by artificial noise introduced by an uncontrolled boundary
correction. One would therefore hope that the samples
used for any such test should be `useful' in the sense defined above
and this, in turn, requires that samples be large enough so as not
to be dominated by boundary effects,
whether the boundary be a simple cube or a more complicated
shape, in much the same way as they
need to be large in order not to be greatly affected by changes
in orientation as discussed above. We shall discuss these
practical issues further, in Section 3 below.

Notice that this definition does not require the construction of a
continuous field on a lattice. One way of treating galaxy data,
for example, is simply to ``decorate'' each galaxy with a sphere
(a sphere is a basic) so that the excursion set is the union of
such spheres. One can then study the behaviour of $\phi_{_H}$ as
a function of the radius of the sphere. This approach, which is reminiscent
of that used in traditional percolation analysis
(e.g. Zel'dovich, Einasto \& Shandarin 1982), is described in detail in
Mecke, Buchert \& Wagner (1994) and Schmalzing et al. (1996),
and also mentioned below.

On the other hand,
one often has to deal with some discretised version of $X$,
perhaps defined on a Cartesian lattice, and consequently
one has a discretised version of the excursion set in which one
cannot locate exactly in $\Re^{3}$
those 
points where $\phi_{_H}({\cal A}\cap {\cal E}_x)$ changes. This would appear to pose some technical
problems. Fortunately, there is a straightforward
implementation of this method to sets defined on regular
lattices. 

A two--dimensional implementation of this idea
is found in Adler (1981); it was introduced to
cosmology by Coles (1988) in the context of regions 
of high or low temperature on the microwave background sky,
and is illustrated in Figure 2. The trick is simply to regard 
each ``basic'' making up the
excursion set as being approximated by the lattice itself.
In 2D, therefore, the basics are approximated as collections
of squares, lines and points. In 3D, we have also cubes. By
simply counting these components one can easily derive
an approximation to the characteristic defined by
equation (\ref{eq:IG1}), which is accurate as
long as the lattice has a finer grid than the typical size
of pieces of the excursion set. The formal proof is
given in Adler (1981, p119); some practical issues are
discussed in Davies (1994). If the excursion set is 
approximated as $S$ squares,
$H$ horizontal lines, $V$ vertical lines and $P$ points then
\begin{equation}
\phi_{_H}= S-[V+H]+P.
\label{eq:h2d}
\end{equation}
Points are counted regardless of whether they belong to lines
or squares; lines are counted regardless of whether they belong
to squares. More detailed investigations of the properties of
this algorithm for two--dimensional
galaxy clustering data are described elsewhere (e.g. Coles \& Plionis 1991;
Plionis, Valdarnini \& Coles 1992; Davies \& Coles 1993;
Coles et al. 1993; Davies 1994).

\begin{figure}
\vspace{10cm}
\caption{The 2D ``lattice'' calculation of the Hadwiger
characteristic. The points inside the contour lie above the
threshold and those outside lie below. Orthogonal lines are
drawn from each point above the threshold to neighbours
which are also above the threshold. In this example the
resulting network has 2 squares, 10 vertical lines, 12 horizontal
lines and 21 points. The characteristic then takes the value
$\chi=2-(10+12)+21=1$, as expected.}
\end{figure}

Corrections for square boundaries can be calculated, where appropriate,
by counting the number of points $P^*$, horizontal lines $H^*$,
and vertical lines $V^*$, in the intersection of the excursion set
with the $x$ and $y$ axes. One then obtains $\chi_{_H}$ according to
\begin{equation}
\chi_{_H}=\phi_{_H}-[P^*-V^*-H^*].
\end{equation}
Methods for dealing with more complicated boundaries are discussed
by Coles \& Plionis (1991), Davies \& Coles (1993) and Davies (1994).

We shall describe the analogous three--dimensional algorithm
for lattice data in Section 3.

\subsection{Comments}
The Euler characterisic $\chi_{_E}$ and the Hadwiger characteristic
$\phi_{_H}$ differ in that they are defined for different classes of
sets and are handled with different kinds of mathematical machinery.
In cases where these two classes overlap, however, the two
definitions are in fact identical (apart from the treatment of
boundaries). Since the sets in which we are interested
actually belong to this overlapping class, we shall henceforth
drop the subscripts and change the notation
to refer to $\chi$ as the Euler--Poincar\'{e}
characteristic, regardless of the way it is measured.
The important point to be made
here is that, although the differential topology approach of section
2.1 is clearly the more elegant route to a 
mathematical derivation of the mean value
of $\chi$ for the excursion sets of a Gaussian random field, that
does not imply that this is necessarly the best way also to 
measure the value of $\chi$
for a given practical realisation of an
excursion set. In fact, it should now be
obvious that this is not
the case: the slicing of
a geometrical body with lines and planes involves much simpler
mathematical concepts and is therefore much easier to implement
on a computer than is the numerical differentiation of 
a fit to the surface of the body.

The algorithm presented in 2.2 gives a much simpler
(and faster) way of extracting $\chi$ for a set of contour data
than integrating the curvature of each contour line around each
compact subset of ${\cal A}_u$. As we shall see, three--dimensional
analyses are simplified even further by the IG approach.
As a practical measuring tool,
therefore, the 3D version of the IG (Hadwiger)
characteristic is generally to be preferred.

\section{The 3D IG Algorithm}

It is first worth re-iterating the point we made above that
the IG approach does not require data
to be presented on a lattice. It is possible to surround each ``galaxy''
in a sample by an appropriately--chosen basic (e.g. a sphere), and then
proceed to slice it in one and two dimensions according to the
prescription of Sec 2.2. This approach has been elaborated by Mecke,
Buchert \& Wagner (1994) and Schmalzing et al. (1996),
 so we will not discuss it further, though it
is an additional clear advantage of the IG approach.

The ``lattice'' algorithm for 3D data is a straightforward generalisation
of the 2D version
described above, and is given mathematical motivation by Adler (1981,
p. 121). In this case, one approximates the shape of
the 3D excursion regions by connecting points above the threshold
into a network of $C$ cubes, $S$ squares, $V$ vertical lines,
$H$ horizontal lines and $P$ points. Points, squares and lines are
counted whether or not they belong to cubes, and so on,
in an analogous manner to
the 2D definition. The appropriate approximation to the Euler--Poincar\'{e}
characteristic is then just
\begin{equation}
\chi = \phi_{_H} = -C +S -[V+H] + P\label{eq:h3d}
\end{equation}
(Adler 1981).
Once again, one has to be a little careful about boundaries. If there is
a simple cubic boundary and one either has a periodic
simulation or is prepared to assume the correction regardless,
then one simply uses the correction
described in Sec. 2.2, namely to subtract off the 2D characteristic
of the intersection of the excursion set with the boundary edges evaluated
using the 2D algorithm. As pointed out above, this means evaluating
three 2D characteristics, one for each of the three faces of the cube
which contain the origin. One can estimate the error in this
correction by simply choosing the origin to be at different corners
of the cube. If these faces contain $P^*$ points, $V^*$ vertical lines,
$H^*$ horizontal lines and $S^*$ squares, then
\begin{equation}
\chi'=\chi_{_H} =\chi-(S^*-[V^*+H^*]+P^*),
\end{equation}
where the prime indicates that a boundary-corrected quantity is being used.
The definition (18) is particularly convenient for simulation data
which are usually presented on such a periodic lattice.

If the boundaries are more complicated than this, such as is almost
certainly the case with a real galaxy sample, then one has to work a little
harder. One approach is one discussed by Coles \& Plionis (1991),
Davies \& Coles (1993) and Davies (1994): one simply takes a 
weighted average the 
characteristic of the excursion set with that of the complementary
set, i.e. the sample volume minus the excursion set; this could be
called the {\em incursion set}. Since a connected region in the
excursion set becomes a hole in the incursion set, the characteristics
of these two sets should be equal (up to a sign), except for the region
around the edge. This method works well in two-dimensions. In a 
separate paper (Pearson et al. 1996), 
we are applying the IG tool to the Abell/ACO cluster
sample and simulations of it based on various models of
cosmological structure formation. Even though this sample is
relatively small and has a complex boundary, 
the correction is small compared to the realisation-to-realisation
variance in the estimator (actually, about 10\%). 

It should be said again that
the DT characteristic does not require any such boundary correction.
Since the results of CONTOUR3D are usually expressed in terms of
$g_S$, which is defined to be related to an integral of the curvature
of the excursion set surfaces.
When one reaches any boundary, therefore,
one simply stops integrating. 
In practical situations, however, this difference is unimportant
because, in order to be useful anyway, a sample should have a
relatively large ratio of interior volume to external boundary.

\begin{figure}
\vspace{15cm}
\caption{Comparison of the results obtained, as a function of $\nu_2$,
for the 3D lattice algorithm (left, labelled ``EPC'') and CONTOUR3D
(right, labelled ``GENUS''). The sign of the EPC curve has been changed
to the $y$-axis scaled by a factor two as indicated by equation (5),
in order to make the results directly comparable. The simulations used 
were of an uncorrelated (i.e. Poisson) Gaussian
random field defined on a $64^{3}$ grid, smoothed on two different
length scales using a Gaussian filter. Results are virtually
indistinguishable. Taken from Davies (1994).}
\end{figure}

Figure 3 shows a comparison of the results of the results obtained
using our IG approach with analogous results obtained using CONTOUR3D
(which, as we have explained, is a product of the DT formalism).
For simplicity, we generated a single 
random-phase realisation of a Gaussian
random fields with a white--noise power spectrum on a $64^{3}$
grid and then analysed it with both algorithms.
The simulations actually have periodic boundary conditions so
we can implement the Adler (1981) correction in the IG
case. We have not inserted error bars on these curves, so
one can see the detailed behaviour of both algorithms more
clearly. Notice that the departures from the Gaussian curve
are similar for both algorithms: on the top simulation,
with a smoothing length of 2.5 grid cells,
the rms difference between
the two different measurements of $\chi$ for this simulation
is of order 0.07 in the units displayed on the $y$-axis. 
This is to be compared with a variation in $\chi$ from
simulation to simulation of order 0.15 determined by
performing an ensemble of simulations of this kind.
This is consistent also with the visual estimate of the scatter
around the ensemble mean. 
Note also that the difference between boundary--corrected
and uncorrected IG estimates is of order 0.03 in this example,
negligible compared to the ensemble variation.
These errors are smaller when the
smoothing is decreased because coherence
length of the field then decreases so that
the excursion set consists of more regions and the
effects discussed previously in connection with the
`fair sample' requirement are strongly reduced. Notice that
the coherence length of the field also determines the
amplitude of the curve: we have scaled this out in these
two examples. 

Of course there are situations where the two methods might give
results that diverge greatly from each other. 
If one had a very large ratio
of coherence length to sample size or a very large grid
spacing, then dealing with the boundary can pose problems
and the method may be inaccurate. On the other hand, even this
may not be too much of a problem in practice. 
Re-iterating a point we
made in Section 2, one is almost always dealing with
a comparison between simulated and observed data sets:
if one is only interested in testing a particular model
then it is less important that a statistic provides an
unbiassed estimate of some globally defined quantity than
that it provides a robust dicriminator between theory and
observation. To a large extent, therefore, one really needs
to ensure only that the same thing is done to both real and
simulated data (same gridding procedure, same boundary, same
smoothing, same correction, etc). A much more detailed
investigation of a confrontation between model and theory
using the IG method is discussed in Pearson et al. (1996).

One should be aware, however, that estimates of $\chi$ made
using the IG method from a  finite sample are not guaranteed
to be unbiassed: one would have to check this carefully using
an ensemble of simulations and adjust any corrections accordingly.
This should warn against using our
method to estimate the coefficient $A$ in equation ({\ref{eq:epgauss})
unless the boundary correction were well--controlled. In any
case, the quantity $\chi$ is not a very efficient estimator
of $A$ so one probably would not wish to employ either the
IG or DT versions of the algorithm to estimate the coherence
length (and hence the power spectrum) anyway. The topology
of the QDOT survey yields only very weak constraints on the
index of the power spectrum of galaxy clustering (Moore et al.
1992). The potential usefulness of $\chi$ lies in its
capability to detect non--Gaussian behaviour whether it
be primordial or induced by non-linear gravitational
evolution of initially Gaussian density perturbations.

\section{Discussion}
In this paper we have introduced an alternative algorithm for
quantifying the topology of excursion sets, with the aim of
producing a simpler tool for measuring the connectivity properties
of high and low density regions in cosmological density fields.

The approach we adopt differs in its mathematical motivation
from the ``traditional'' approach. We have therefore gone to
some length to attempt to explain the foundations of the
different approaches
to topological analyses based on differential topology (DT), on the
one hand, and integral geometry (IG) on the other. The central point to
emerge from these considerations is that the DT approach is better
for handling the properties of random fields analytically (mainly
because one can use the powerful mathematics of Morse theory to
relate integrated curvature to the simple counting of critical points).
The IG approach, although not conducive to analytic studies to anything
like the same extent, nevertheless generates much simpler algorithms
for measuring the relevant characteristics in practical situations.
Importantly (though we did not include the proof here), the situations
in which we are interested in exploiting such characteristics involve
objects which can, in principle, be treated with either approach. The moral
of this tale is, therefore, that DT is better for theory, IG better
for practice.

Although the IG algorithm has the advantage of being faster
(it is also parallelisable), this is probably not a particularly
relevant consideration for present large--scale structure studies
because the computation of the so--called ``genus'' curves is not
in any case an arduous task. Future generations of redshift surveys,
which will be an order of magnitude larger in size than existing ones
may invalidate this argument. Our particular motivation for
favouring this algorithm at the present time, however, has been the
need to analyse the topological properties of a huge ensemble of
realisations of large--scale cluster simulations (Borgani et al. 1995),
for which the speed and simplicity of the new algorithm has great
advantages. This latter study, 
(Pearson et al., in preparation), which also takes into account
percolation properties and measures of ``filamentariness'',
is intended to provide a theoretical understanding of the
differences in clustering pattern for models which differ
only slightly in their initial power spectra and cosmological
parameters. Quite apart from this specific argument, however,
we stress the more obvious point that simpler algorithms
should generally be preferred to complicated ones which
produce the same results.

\section* {Acknowledgments}
 PC is a PPARC Advanced Research Fellow. 
AGD received an SERC postgraduate studentship and RCP a PPARC
postgraduate studentship while this work was being done. We are
all extremely grateful to the referee, Thomas Buchert,
 for his thorough reading
of a previous version of this paper, and for his many perceptive comments.

\end{document}